# A TECHNICAL MEMORANDUM ON CORE RADII IN LENS STATISTICS

CHRISTOPHER S. KOCHANEK

*Harvard-Smithsonian Center for Astrophysics*
*60 Garden Street, Cambridge, MA 02138, USA*

**Abstract.** Quantitative estimates of lensing probabilities must be self-consistent. In particular, for asymptotically isothermal models: (1) using the $(3/2)^{1/2}$ correction for the velocity dispersion overestimates the expected number of lenses by 150% and their average separations by 50%, thereby introducing large cosmological errors; (2) when a core radius is added to the SIS model, the velocity dispersion must be increased; and (3) cross sections and magnification bias cannot be separated when computing the lensing probability. When we self-consistently calculate the effects of finite core radii in flat cosmological models, we find that the cosmological limits are independent of the core radius.

## 1. Introduction

Asymptotically isothermal potentials are consistent with most data on gravitational lenses. They explain the observed numbers of lenses (Maoz & Rix 1993, Kochanek 1993), fit most observed image configurations (e.g. Kochanek 1991a), and are consistent with stellar dynamics (Kochanek 1994, Franx 1993). Photometry of early type galaxies (Tremaine et al. 1994), and the absence of central images in most lenses (Wallington & Narayan 1993, Kassiola & Kovner 1993) suggest that the lens potentials have a small or vanishing core radius. There is, however, a persistent myth (generated in part by the author) that small core radii can dramatically alter the expected number of lenses without other observational consequences. We can trace the current versions of this myth to inconsistencies in either the dynamical normalization of the models or the calculation of the magnification bias. In this technical memorandum we briefly explore these consistency re-quirements and the resulting effects of a finite core radius on cosmological limits.

We confine ourselves to a circular isothermal density distribution with $\rho = \sigma_{DM}^2/2\pi G(r^2 + s^2)$ where $\sigma_{DM}$ is the velocity dispersion of the dark matter, and $s$ is the core radius (Hinshaw & Krauss 1987). The lens deflects rays by

$$\frac{\partial \phi}{\partial r} = b \frac{(r^2 + s^2)^{1/2} - s}{r} \qquad (1)$$

where $b = 4\pi(\sigma_{DM}/c)^2 D_{LS}/D_{OS}$, $D_{LS}$ and $D_{OS}$ are proper motion or angular diameter distances between the lens and the source and the observer and the source respectively, and $\phi$ is the two-dimensional lensing potential. (The alternate softened isothermal model with $\phi = b(r^2 + s^2)^{1/2}$ introduced by Blandford & Kochanek (1987) is almost indistinguishable from the Hinshaw & Krauss (1987) model if its core radius is twice as large.)

The lens is supercritical and able to generate multiple images if $\beta = s/b < 1/2$, with a tangential critical line at $r_+ = b(1 - 2\beta)^{1/2}$ and a radial critical line at $r_- = b[\beta - \beta^2/2 - \beta^{3/2}(4 + \beta)^{1/2}]^{1/2}$. The caustics are at $u_+ = 0$ and $u_-$ where the cross section is $\tau = \pi u_-^2 = \pi b^2[1 + 5\beta - \beta^2/2 - \beta^{1/2}(4 + \beta)^{3/2}/2]$ (Hinshaw & Krauss 1987). The cross section declines very rapidly with $\beta$ and near the threshold of $\beta = 1/2 - \epsilon$ it declines as $\tau \propto \epsilon^3$ if $b$ is held fixed. If we assume a *constant comoving core radius* the cross section can be integrated analytically to compute the optical depth (Krauss & White 1992).

## 2. Dynamical Normalizations

The first question we must address is the normalization of the singular model ($s \to 0$). Historically, Turner, Ostriker, & Gott (1984) argued that if the central velocity dispersion of the stars is $\sigma_c$ then the dark matter should have velocity dispersion $\sigma_{DM} = (3/2)^{1/2}\sigma_c$. However, Franx (1993), Kochanek (1993, 1994), and Breimer and Sanders (1993) show convincingly that real galaxies do not satisfy the assumptions used by Turner, Ostriker, & Gott (1984), and that for real galaxies $\sigma_{DM} \simeq \sigma_c$. Kochanek (1994) fit a sample of 37 early type galaxies from van der Marel (1991) and found that the best fit estimate was $\sigma_{DM*} = 225 \pm 10$ km s$^{-1}$ for an $L_*$ galaxy.

All existing studies of the effects of a core radius on lens statistics have added a core radius while leaving the velocity dispersion $\sigma_{DM}$ or $b$ unchanged. It is clear, however, that $s$ and $\sigma_{DM}$ must be correlated. Adding a core radius reduces the mass near the center of the galaxy, and the velocity dispersion must increase compared to its value in a singular model to maintain either the stellar velocity dispersions or the average image separations fixed. As a model calculation, we compute the average line-of-sight veloc-ity dispersion inside one effective radius $R_e$ assuming a Hernquist (1990) distribution ($\nu \propto r^{-1}(r + a)^{-3}$) for the stars with $a \simeq 0.45 R_e$. With the assumption that the velocity dispersion tensor of the stars is isotropic, the dark matter dispersion increases as $\sigma_{DM} \propto 1 + 2(s/R_e)$ with the addition of a core radius.

For a more realistic model we use van der Marel (1991) sample and fit isotropic dynamical models to each galaxy, assuming the core radius is a constant fraction of the estimated effective radius for each galaxy, and that the velocity dispersion scales as $L/L_* = (\sigma_{DM}/\sigma_{DM*})^4$. The $\chi^2$ surface of the fit to the observed velocity dispersion profiles is shown in Figure 1, and the dashed line is the scaling law estimated from the Hernquist (1990) model. Models with large core radii cannot fit the data because of the contradiction between a homogeneous core and a steeply rising luminosity profile. The formal 95% confidence upper limit on the core radius is $s_* \lesssim 0.08 R_e$ or $s_* \lesssim 300 h^{-1}$ pc for $R_{e*} = 4 h^{-1}$ kpc. For a core radius of $s_* = 100 h^{-1}$ pc the fractional increase in the velocity dispersion is 7.5% or 17 km s$^{-1}$, less than the uncertainty in the value of $\sigma_{DM*}$. Nonetheless, its effects on models with a finite core radius are striking; it produces a 33% increase in the expected number of lenses if we keep the ratio of the core radius to the critical radius fixed ($s/b$ constant).

Self-consistency in lens models also requires a velocity dispersion that increases as the core radius becomes larger since the average image separation must stay fixed as the core radius increases. The image separation is approximately twice the tangential critical radius of the lens ($\Delta\theta \simeq 2r_+ = 2b(1 - 2\beta)$), so that if the core radius increases (larger $\beta$), the only way to maintain constant average image separations is to also increase the average velocity dispersion (larger $b$). If we model this by keeping the tangential critical radius $r_+ = b(1 - 2\beta)$ fixed, then the lens cross section $\tau$ decreases as $\tau \propto \epsilon^2$ instead of $\epsilon^3$.

## 3.  Magnification Bias

Self-consistent calculations of the lensing probability such as Kochanek & Blandford (1987), Kochanek (1991b, 1993), Wallington & Narayan (1993), Kassiola & Kovner (1993), and (in most regimes) Maoz & Rix (1993) automatically include the effects of the core radius on the magnification bias, but most treatments of softened isothermal models examined only the effects of a core radius on lensing cross sections (e.g. Dyer 1984, Blandford & Kochanek 1987, Hinshaw & Krauss 1987, Krauss & White 1992, Fukugita & Turner 1991, Fukugita et al. 1992, Bloomfield-Torres & Waga 1995). Core radii have a powerful effect on the cross section for multiple imaging. However, using the change in the cross section grossly overestimates the effects of a finite core radius on the lensing probability in bright quasar samples. The core radius first eliminates images with low total magnifications, but the bright quasar samples are dominated by highly magnified images of fainter quasars and magnification bias significantly reduces the effects of adding a core radius on the probability.

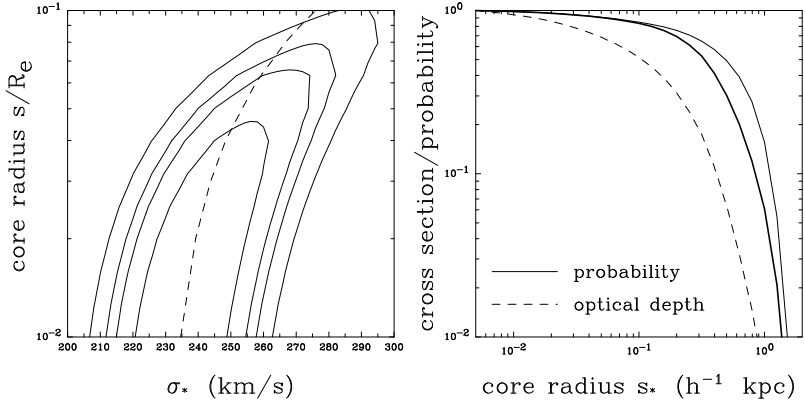

*Figure 1.* (Left) Contours of the $\chi^2$ for the dynamical fits. The light solid lines show the 68%, 90%, 95%, and 99% confidence limit changes on $\Delta\chi^2$ for the fit to the sample. The dashed line shows the expected scaling of $\sigma_{DM*}$ with $s/R_e$ if we keep the average velocity dispersion interior to $R_e$ fixed in the Hernquist/softened isothermal sphere dynamical model.

*Figure 2.* (Right) Variation in cross section (dashed line) and lensing probability (solid line) with core radius $s$ for a lens with $\sigma_{DM} = 250$ km s$^{-1}$. The values are normalized to unity at the minimum core radius. The heavy solid line shows the lensing probability excluding image systems with detectable central images using the same normalization as for the total probability. The results are given for the average over the quasar data sample including selection effects.

We can understand this analytically in the Hinshaw & Krauss (1987) model. Inconsistent models of the effects of a core radius estimate the lensing probability by using the optical depth multiplied by the magnification bias for the singular model. When two images are merging on the radial caustic, the third image is located at $r_{out} = 2\beta u_-/r_-^2$, and the average magnification produced by the lens is $\langle M \rangle = r_{out}^2/u_-^2 = 4\beta^2/r_-^4$. When the core radius is small, the average magnification is 4, but near the threshold the average magnification diverges as $\langle M \rangle \propto \epsilon^{-2}$. The magnification probability distribution is approximately $P(> M) = (\langle M \rangle/2M)^2$ when $M > \langle M \rangle/2$ for fold caustic statistics (e.g. Blandford & Narayan 1986). If we assume a single power law quasar number counts distribution with $dN/dm \propto 10^{\gamma(m-m_0)}$ then the magnification bias varies with the average magnification as $B(m) \propto \langle M \rangle_0^{2.5\gamma}$ for $\gamma < 0.8$. As the core shrinks, the average magnification increases, which drives up the magnification bias. The lensing probability, including the change in the magnification bias, varies as $\tau B(m) \propto \epsilon^{3-5\gamma}$ not $\tau \propto \epsilon^3$. For large average magnifications the effective value of $\gamma$ is the faint slope of the quasar number counts, $\gamma \simeq 0.27 \pm 0.07$, and $\tau B(m) \propto \epsilon^{1.65 \pm 0.35}$. For bright quasars the increase in the bias is greater because of the steeper number counts slope, and the effects of the core radius are still smaller.

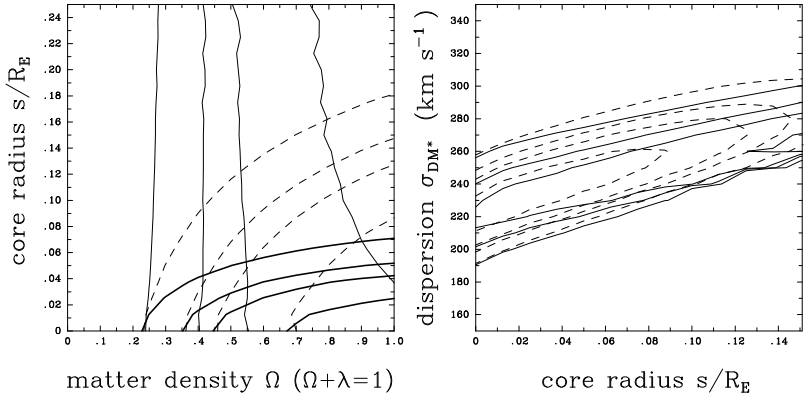

*Figure 3.* Cosmological effects of finite core radii. The left panel shows contours as a function of the ratio of the core radius to the effective radius and the cosmological model. The light solid lines are the constraints from lensing alone, the dashed lines adds the velocity dispersion prior as a function of core radius estimated in §2, and the heavy solid lines add the goodness of fit to the dynamical data. The right panel shows dependence of $\sigma_{DM*}$ on the core radius in an $\Omega_0 = 1$ cosmology. The light contours use only the lens data, and the dashed lines include the prior probability distribution for $\sigma_{DM*}$ estimated from dynamical models. The contours are drawn at the 68%, 90%, 95%, and 99% confidence levels for one parameter.

To emphasize this point, Figure 2 shows the relative variation of the cross section and the true lensing probability including magnification bias for a lens with $\sigma_{DM} = 250$ km s$^{-1}$ as a function of the core radius averaged over the full quasar data sample. For a core radius of $s = 100h^{-1}$ pc using the cross section instead of the true probability underestimates the lensing probability by about 40%. This comparison overestimates the effect of a finite core radius because we did not include the dependence of $\sigma_{DM}$ on $s$.

## 4. The Cosmological Effects of Softened Isothermal Spheres

We assume that the core radii of galaxies are proportional to their effective radii $s = s_*(L/L_*)^{1.2}$, and the models are characterized by a fixed ratio of $s/R_e$. To simplify the calculations, we set the "Faber-Jackson" exponent to be $\gamma = 4$, so the core radius varies with the velocity dispersion as $s = s_*(\sigma_{DM}/\sigma_{DM*})^{4.8} = s_*(L/L_*)^{1.2}$ (see Kochanek 1995 for details).

The right panel of Figure 3 shows the best fit value of $\sigma_{DM*}$ as a function of $s_*$ in an $\Omega_0 = 1$ cosmology. As expected from §2, the velocity dispersion increases as the core radius increases with $\sigma_{DM*} \propto 1 + s/R_e$. This is shallower than the slope seen in the dynamical models. The left panel of Figure 3 shows the dependence of the cosmological limits for a flat universe ($\Omega_0 + \lambda_0 = 1$) on the core radius using only the lens data, the lens data combined with the prior probability distribution for $\sigma_{DM*}$ derived from the dynamical model of §2, and finally the lens data, the dynamical velocity dispersion prior, and the likelihood of the dynamical model. If we use only the lensing data, the cosmological limits are nearly independent of the core radius – this is a radically different picture of the effect of a core radius than that found in inconsistent calculations.

## References

Blandford, R.D., & Narayan, R. 1986, ApJ, 310, 568

Blandford, R.D., & Kochanek, C.S., 1987, ApJ, 321, 658

Bloomfield-Torres, L.F.B., & Waga, I., 1995, this volume

Breimer, T.G., & Sanders, R.H., 1993, A&A, 274, 96

Dyer, C.C., 1984, ApJ, 287, 26

Franx, M., 1993, in Galactic Bulges, ed. H. Dejonghe & H.J. Habing, (Dordrecht: Kluwer) 243

Fukugita, M., & Turner, E.L., 1991, MNRAS, 253, 99

Fukugita, M., Futamase, T., Kasai, M., & Turner, E.L., 1992, MNRAS, 393, 3

Hernquist, L., 1990, ApJ, 356, 359

Hinshaw, G., & Krauss, L.M., 1987, ApJ, 320, 468

Kassiola, A., & Kovner, I., 1993, ApJ, 417, 450

Kochanek, C.S., 1991a, ApJ, 373, 354

Kochanek, C.S., 1991b, ApJ, 379, 517

Kochanek, C.S., 1993, ApJ, 419, 12

Kochanek, C.S., 1994, ApJ, 436, 56

Kochanek, C.S., 1995, submitted to ApJ

Kochanek, C.S., & Blandford, R.D., 1987, ApJ, 321, 676

Krauss, L.M., & White, M., 1992, ApJ, 394, 385

Maoz, D., & Rix, H.-W., 1993, ApJ, 416, 425

Tremaine, S., Richstone, D.O., Yong-Ik Byun, Dressler, A., Faber, S.M., Grillmair, C., Kormendy, J., & Lauer, T.R., 1994, AJ, 107, 634

Turner, E.L., Ostriker, J.P., & Gott, J.R., 1984, ApJ, 284, 1

van der Marel, R.P., 1991, MNRAS, 253, 710

Wallington, S., & Narayan, R., 1993, ApJ, 403, 517